 \documentclass[journal,twocolumn]{IEEEtran}
\usepackage{graphicx}
\usepackage{comment}
\usepackage{amssymb}
\usepackage [noadjust]{cite} 
\usepackage{bm}  
\usepackage{amsmath}
\usepackage{amssymb}
\usepackage{enumerate}
\usepackage{stfloats}
\usepackage{cases}
\usepackage{epstopdf}
\usepackage{soul,color} 
\usepackage[percent]{overpic}
\usepackage{tabularx}
\usepackage{subfigure}
\usepackage[table]{xcolor}
\usepackage{multirow}
\usepackage{booktabs}  
\usepackage{tabu} 
\usepackage{soul}

\usepackage{algorithm}
\usepackage{algorithmic}
\usepackage{todonotes}

 \newcommand{\algstartblock}
 {	
 	\ifnum\value{algcounter}<10{
 		\addtocounter{algblockspace}{12}
 	}\else{
 		\addtocounter{algblockspace}{12}
 	}
 	\fi
 }
 \newcommand{\algendblock}{
 	\ifnum\value{algcounter}<10{
 		\addtocounter{algblockspace}{-12}
 	}\else{
 		\addtocounter{algblockspace}{-12}
 	}
 	\fi
 }

 \newcommand{\alg}[1]{
	\ifnum\value{algcounter}<10{
	 	\begin{enumerate}[\arabic{algcounter}.\hspace{\numexpr\value{algblockspace}+0 pt}]
	 		\setlength\itemindent{-7pt}
			\item \hspace{-8pt} #1
		\end{enumerate} \stepcounter{algcounter}
	}\else{
		\begin{enumerate}[\arabic{algcounter}.\hspace{\numexpr\value{algblockspace}+0 pt}]
			\setlength\itemindent{-11pt}
			\item \hspace{-8pt} #1
		\end{enumerate} \stepcounter{algcounter}
	}
	\fi
  }

 \newcommand{\Rdddirectone}[3]{R^{dir_1}}
 \newcommand{\Rdddirectzero}[2]{R^{dir_0}}

 \newcommand{\Rddrelay}[2]{R^{rel}}

 \newcommand{\Rcellularzero}[2]{R^{cel_0}}
 \newcommand{\Rcellularone}[3]{R^{cel_1}}

\begin{document}
\raggedbottom
\allowdisplaybreaks
\title{User Association in Coexisting RF and TeraHertz Networks in 6G}
\author{Noha~Hassan, Md Tanvir Hossan, {\em  Member IEEE}, and~Hina~Tabassum, {\em Senior Member IEEE}
     \thanks{This work is supported by the Discovery
Grant from the Natural Sciences and Engineering Research Council of Canada. 
Noha Hassan was a visiting researcher at  the Department of Electrical Engineering and Computer Science,  York University, Canada and now is affiliated with the Department of Electrical and Computer Engineering, Ryerson University, e-mail: noha.hassan@ryerson.ca. Md Tanvir Hossan and H. Tabassum are with the Department of Electrical Engineering and Computer Science,  York University, e-mail: mthossan@yorku.ca, hinat@yorku.ca.}
}
\date{} 

\maketitle

\begin{abstract}
While fifth generation (5G) networks are ready for
deployment, discussions over sixth generation (6G)  networks are down the road. 
Since  high frequencies like terahertz (THz) will be central to 6G, in this paper,   we propose two user association (UE) algorithms considering a coexisting RF and THz network that balances the traffic load across the network by minimizing the standard deviation of the network traffic load. 
 Our algorithms capture the  heterogeneity observed at RF and THz frequencies such as transmission bandwidth, molecular absorption, transmit powers, etc. Unlike typical unsupervised clustering algorithms (e.g. k-means, k-medoid, etc.) that search for appropriate cluster centers' locations, our algorithms identify the appropriate UEs to be associated to a certain BS such that the overall network load standard deviation (STD) can be minimized subject to users' rate constraints. In particular, our algorithms cluster UEs to every base station (BS) such that the traffic load across the network can be balanced, i.e., by minimizing the STD of network traffic load. Numerical  results  show  that  the proposed algorithms outperform the classical user association algorithms in terms of  data  rate,  traffic  load  balancing,  and  user’s  fairness.

\end{abstract}

\raggedbottom
%
\section{Introduction}

With the recent release of  5G-new radio (NR) standard, early 5G deployment has already started in various countries including South Korea, Canada, and China. 
5G-NR has a multitude of advantages over the long-term evolution (LTE)/LTE-advanced technology, i.e., higher data rates ($\sim$ 0.1~Gbps), low latency  ($\sim$ 1 - 10~msec), higher mobility ($\sim$ 500~km/h), and support to 10$^6$ devices per sq. km. 
The intriguing use cases of 5G-NR, such as ultra-reliable low latency communication (uRLLC), enhanced mobile broadband (eMBB), and massive machine-to-machine communication (mMTC) leverage on three disruptive technologies, i.e., millimeter-wave (mm-wave) communication, large-scale antenna arrays (i.e., massive MIMO), and ultra-dense deployment of access points. 

Despite the aforementioned advancements, the  global  mobile  traffic volume is expected to grow from  7.462~EB/month  in  2010 to  5016  EB/month  in 2030  \cite{8631208}. Thus,  the  launch  of  sixth  generation  (6G) wireless  networks is inevitable.  6G networks will observe coexisting RF and mm-wave deployments \cite{ibrahim2019meta} and coexisting RF and visible light communication (VLC) \cite{tabassum2018coverage, elgala2019introduction} deployments. In addition, higher frequencies in the terahertz (THz) band [0.1-10 THz] will be central to ubiquitous wireless communications in 6G. 
THz frequencies promise to support ample spectrum, above hundred Gigabit-per-second (Gbps) data rates, massive connectivity, denser networks, and highly secure transmissions. Multiple leading 6G initiatives probe THz communications, including the ``6Genesis Flagship Program (6GFP)", the European Commission’s H2020 ICT-09 THz Project Cluster, and the “Broadband Communications and New Networks" in China. In the US, THz technology was identified in 2014 by the US Defense Advanced Research Projects Agency (DARPA) as one of the four major research areas that could  impact  society more than the internet. Similarly, the US National Science Foundation and the Semiconductor Research Consortium (SRC) also identify THz as one of the four essential components of the next IT revolution.


The THz spectrum exists between the mmWave and the far-infrared (IR) bands and has, for long, been the least investigated electromagnetic spectrum. However, recent advancements in THz signal generation, modulation, and radiation methods are closing the so-called THz gap. Nevertheless, channel propagation at THz frequency bands is susceptible to molecular absorption, blockages, atmospheric gaseous losses due to oxygen molecule and water vapor absorption. On the other hand, conventional RF spectrum is characterized with strong transmission powers and wider coverage; however, the spectrum is limited and extremely congested. 

Evidently, THz  networks have reduced coverage  but plenty of spectrum, so there exists a trade-off based on users’  channel quality and available spectrum. 
\textit{To overcome the trade-offs between different frequencies, opportunistic spectrum selection mechanisms should be designed considering a coexisting  network where RF BSs and THz BSs coexist.} The first work that considered a coexisting RF and dense THz network was presented in \cite{sayehvand2020interference}. 
In a coexisting network, due to the enhanced signal power from RF BSs, it is highly likely that the user will be biased towards RF SBSs, albeit the THz BSs can provide very large transmission bandwidth yielding very high data rates and ultra-low latencies. 
As such, new traffic offloading and user clustering schemes will be crucial where users can be offloaded to different BSs and the resource utilization can be improved by balancing the traffic load among BSs.

In this paper,   we propose two user equipment (UE) association  algorithms based on unsupervised learning in a coexisting RF/THz  network. Our algorithm captures the network heterogeneity observed at RF and THz frequencies. The algorithms cluster UEs to every base station (BS) such that the traffic load across the network can be balanced, i.e., by minimizing the standard deviation of network traffic load. Numerical  results  show  that  the proposed algorithms outperform the classical algorithms in terms of  data  rate,  traffic  load  balancing,  and  user’s  fairness.
Unlike typical unsupervised clustering algorithms (e.g. k-means, k-medoid, etc.) that search for appropriate cluster centers' locations, our algorithm identifies the appropriate UEs to be associated to a certain BS such that the overall network load standard deviation (STD) can be minimized subject to rate constraints. We use standard deviation as a measure to identify how many BSs are overloaded and underloaded and need to redistribute UEs among them. Standard deviation depends on distribution of UEs at every BS, so when distribution changes standard deviation will change as well.  We consider every BS as an independent cluster and associate UEs to each BS. STD minimization enables traffic load balancing among BSs and reduces load variations in every BS  from the mean value. 

\section{State-of-the-art: User Association}
To date, several user association methods have been explored for 5G heterogeneous networks (HetNets) with the objective of traffic load balancing (a survey is provided in \cite{liu2016user}). Some popular methods include cell-range expansion (CRE) \cite{muhammad2016cell} and resource-aware UE association algorithms \cite{hossain2014evolution}. 

In \cite{muhammad2016cell}, the authors combined UE association algorithms using SBS along with traffic offloading using CRE.  The drawback of CRE is that the MBS acts as a strong interferer and almost-blank subframes based strategies are required. In \cite{hossain2014evolution}, a resource aware user association scheme has been proposed. In \cite{alizadeh2019load}, the authors proposed a new user association algorithm that considers the highly directional mm Waves, network interference, and their vulnerability to small channel variations. The proposed algorithm is dependent on the network interference structure and user association adjusting the interference according to the association, and under the max-min fairness.
Also, in \cite{liu2019joint}, the authors considered the spectrum heterogeneity of mmWave frequency bands by introducing two access schemes; the single-band and multi-band access schemes. For the first access scheme, the authors developed an iterative algorithm  based on the Lagrangian dual decomposition methods and the Newton-Raphson method for joint user association and power allocation. For the multi-band access scheme, a Markov approximation framework is used to develop a near-optimum user association algorithm. The results revealed that different users can only access  one band simultaneously. In  \cite{xu2018adaptive}, the authors solved a mixed-integer optimization to maximize the network throughput of time-variant mmWave networks, and suggested that distributed association techniques will solve the problems of wireless channel variations (due to obstacles) and client mobilities. 

For THz-only network, \cite{boulogeorgos2018users} introduced a user association algorithm to maximize the total throughput that takes into account the directivity and position of the BSs’ and UEs’ antennas besides the minimum rate requirements using the grey wolf optimizer. The proposed framework proved to be more efficient than the commonly used particle swarm optimizer (PSO) approach and that the only required control parameters are the population size and number of iterations.

In \cite{ye2013user}, the authors devised an algorithm to increase the load for the lightly loaded small cell BSs. They solved a logarithmic utility maximization problem considering multi-association to BSs, where the equal resource allocation converges to near-optimal solution. In \cite{abbas2017analysis}, the authors employed a selective method of UE association, where the MBS coverage is divided into center and edge regions and SBSs are only active in the edges. In \cite{2004.07469} authors derived connection probability and the average ergodic capacity for two types of multi-connectivity, such as closest line-of-sight access point and reactive connectivity. One of the important analysis from their model is that authors are taking the blockage into account for user association problem. However, there are plenty of BS that can be present in the same territory, which is  ignored. 

\section{Network Model}
The conventional RF macro base-stations (MBSs) and THz base-stations (TBSs) are randomly deployed. MBSs and TBSs are equipped with $M_R$ and $M_T$ antennas with {$S_R$ and $S_T$ orthogonal} streams. Total number of MBSs and TBSs is { $B$}. The users are randomly deployed and their total number is $U$. From Eq. 2 in \cite{7456247}, we can determine data rate for a user associated to tier $k \in \{R,T\}$.
{$
\mathrm{R}_k = W_k \log_{2} \left(1+ \frac{M_k-S_k+1}{S_k}\mathrm{SINR}_k\right),  k \in \{R,T\}
$
where $W_k$ denotes the bandwidth available at tier $k$. The massive MIMO regime refers to as the case where $1<<S_k<<M_k$. The factor in the numerator $(M_k-S_k+1)$ is the massive MIMO gain at the user.} 
In what follows, we describe the channel propagation and signal-to-interference-plus-noise ratio (SINR) at THz and RF frequencies. 

\subsubsection{RF Channel and SINR Model}  The RF channel experiences both the channel fading  and path-loss. Thus, the received signal power at the typical user can be modeled as $h(\rho)=\gamma_{R} \rho^{-\alpha} \chi$, where $\chi$ is the exponentially distributed (Rayleigh fading) channel power with unit mean from the tagged SBS, {$\alpha$} is the path-loss exponent, and $\rho$ is the distance of the considered user to the serving SBS. Also, $\gamma_{R} = \frac{c^2}{\left(4\pi f_{R}\right)^2}$, where  $f_{R}=300$ MHz is the RF carrier frequency, and $c = 3 \times 10^8$ m/s is the speed of light. Based on this, the SINR of a typical user on RF  channel can be calculated {as $\frac{M_R-S_R+1}{S_R} \frac{P_{R}\gamma_{R}\rho_0^{-\alpha}\chi_0}{N^R_{0}+I_{\mathrm{agg}}^{R}}$, where { $P_{R}$ is the transmit power of the SBSs and $N^R_{0}$ }is the thermal noise at the receiver.} 



\subsubsection{THz Channel and SINR Model}  Since the molecular absorption loss is high in THz, the impact of multi-path fading and NLOS transmission is negligible. Thus,  we model the LOS channel power between users and TBSs  as
$
h \left(\rho\right) =  \gamma_T{\rho^{-2}{\exp\left(-k_{a}\left(f\right)\rho\right)}},$ 
where $\gamma_T=\frac{c^2}{\left(4\pi f_{T}\right)^2}$, $k_{a}\left(f\right)$ is the molecular absorption coefficient, $r$ is the distance between the transmitter and  receiver, $f_{T}$ is the frequency at which the THz devices are operating, and $c$ is the speed of light. The LOS SINR of the typical user associated to its desired TBS can be calculated as
{$
\frac{M_T-S_T+1}{S_T} \frac{P_{\mathrm{T}} \gamma_T \rho_0^{-2} {\exp\left(-k_{a}\left(f\right)\rho_{0}\right)}}{N^T_{0}+I_{\mathrm{agg}}^{T}},
$}
where $P_{T}$ is the transmit power of the TBS and 
{ $N^T_{0}=K_{B} T + P_{T_{j}} \gamma_{T_{j}} \rho_{j}^{-2}(1-e^{-k_a(f)\rho_{j}})$,} noise power denotes the  thermal noise and the noise resulted from the molecular absorption. 

\section{Proposed Algorithms}
In this section, we propose two unsupervised clustering algorithms which are different from conventional unsupervised clustering algorithms referred to as {\em Least Standard Deviation}-based clustering and {\em mean traffic load}-based clustering algorithms. For example, K-means clustering uses square of the distance from a centroid to minimize the clustering error. Other modified K-means methods depend on standard deviation where they search for the location of the cluster having the maximum standard deviation from a centroid \cite{thangavel2006combined, inproceedings}. 
\subsubsection{Least Standard Deviation User Clustering Algorithm}

In our algorithm, a binary matrix $l$ is generated by choosing an acceptable level of SINR for UEs and discarding the  UEs (by assigning a logical value of 0) who cannot connect to the corresponding BSs and have less SINR levels. If a UE has acceptable SINR value from a BS, the UE assignment takes the logical value of 1. We define number of possible BSs per UE, and start from a UE with the least possible available BSs, where UE is assigned if there is only one possible BS. If UE has more than one possible BS, then choose the BS with least possibilities first and attempt connection to the least loaded BS. For other UEs in the network, consider the least loaded BS then add number of UEs associated to every BS. Calculate load STD of the network and check whether if it is less than a certain threshold.
Repeat for next UE with least available BSs until algorithm converges and network load STD is less than a chosen threshold. 
The procedure is detailed in {\bf Algorithm~1} and in the steps below, (i)~Initialize number of BSs, number of UEs, SINR threshold, load per BS vector,~and  standard deviation threshold is 1, (ii)~Calculate SINR matrix and generate binary matrix $l$, (iii)~Compute number of BSs per UE $nbs_{u}$ and  number of  UEs  per BS $nu_{bs}$, (iv)~Start from UE with smallest $nbs_{u}$, (v)~{If} {$nbs_{u}=1$ {then}} associate UE to BS, otherwise attempt connecting the UE to a BS with smallest $nu_{bs}$, (vi)~Calculate final load per BS { L} and {\bf STD}
(vii)~{If} {STD$ < \epsilon$ {then}} optimum load per BS obtained, otherwise repeat the steps {until} STD$ > \epsilon$.

\begin{algorithm}[t]
 \caption{Least Standard Deviation (LSTD)}
 \begin{algorithmic}[1]
 \STATE \textbf{Initialize:} $B$: No. of BSs,~$U$: No. of UEs, $\gamma$: SINR Threshold, \textbf{L}: load per BS vector=0,~and $\epsilon =1$.
  \STATE \textbf{while} $STD > \epsilon$  \textbf{do}\\
  \STATE \quad Calculate SINR Matrix\\
    \STATE \quad Generate binary matrix $l$ \\
      \STATE \quad Compute number of BSs per UE $nbs_{u}$ 
      \STATE \quad Compute  number of  UEs  per BS $nu_{bs}$\\
      \STATE \quad Start from UE with smallest $nbs_{u}$\\
\STATE \quad\textbf{If} {$nbs_{u}(B)=1$ \textbf{then}}\\ 
\STATE \qquad Associate UE to BS\\
\STATE \quad {\bf else } 
\STATE \qquad Attempt connecting to smallest $nu_{bs}$\\
\STATE \quad Calculate final load per BS \textbf{L} and {\rm STD}\\
\STATE \textbf{If} {$STD < \epsilon$ \textbf{then}}\\
\STATE \quad Optimum \textbf{L} obtained\\
\STATE \textbf{Repeat}\\
\end{algorithmic}
\end{algorithm}

\subsubsection{Redistribution of BSs Load (RBL)-based Clustering}
We propose a non-parametric unsupervised learning algorithm.
Our algorithm forms BS clusters based on UEs' calculated SINR levels and not according to Euclidean distance, as signal strength is the main concern. First, UEs are associated to BS clusters based on maximum SINR value. Our algorithm learns from the load per BS and defines BS status (that some BSs might be overloaded and some are under loaded). Re-association of UEs is carried out where over loaded BS clusters lay off some of their UEs (The UEs with the strongest signals are chosen) and donate them to BSs with less load. As UEs getting the strongest signals are chosen, then signal quality will not be affected when load balancing is carried out.
Our algorithm is real-time, where the load can be redistributed instantly from one cluster to another. Mean value of UEs per BS ($\mu$) varies for every tier as every BS has a certain capacity to associate UEs (due to different transmission bandwidth available in RF and THz). Mean value is defined as the maximum number of UEs that can be associated to a BS (considering UEs' traffic demand) divided by number of BSs in a tier.

The procedure is detailed in {\bf Algorithm~2} and in the steps below, (i)~Initialize number of BSs $B$, number of UEs $U$, mean value of UEs per BS $\mu$, number of UEs per BS $N_{u}$. (ii)~Calculate SINR Matrix and associate UEs to BSs based on Max-SINR. (iii) Calculate $\mu$ per tier and $N_{u}$. {If} $N_{u} > \mu$ {then} BS status is "overloaded". (iv) Sort UEs associated to "overloaded" BSs based on highest SINR. {If} $N_{u}< \mu$ {then}  BS status is "accepting". (v) Sort accepting BSs based on highest SINR value.  Move UE on top of the "overloaded" BSs list to BS on top of the "accepting" BSs list.  Repeat for rest of UEs of first "overloaded" BS. Repeat for rest of "overloaded" BSs. (vi) {If} $N_{u} =\mu$ {then}  finalize UEs associated with that BS to it. For moving or newly added UEs, calculate the new SINR matrix and repeat. Repeat for the other tier (TeraHertz) and generate final load distribution per BS.
\begin{algorithm}[t]
 \caption{Redistribution of BSs Load (RBL)}
 \begin{algorithmic}[1]
 \STATE \textbf{Initialize:} $B$: No. of BSs,~$U$: No. of UEs,~$\mu$: Mean value of UEs per BS,~$N_{u}$: No. of UEs per BS\\
  \STATE Calculate SINR Matrix\\
    \STATE Associate UEs to BSs based on Max-SINR\\
    \STATE Calculate $\mu$ per tier and $N_{u}$\\
     \STATE \textbf{If} $N_{u} > \mu$ \textbf{then}\\ 
    \STATE BS status is "overloaded"\\
     \STATE Sort UEs associated to "overloaded" BS based on highest SINR.
     \STATE \textbf{If} $N_{u}< \mu$ \textbf{then}\\ 
      \STATE BS status is "accepting"\\
      \STATE Sort accepting BSs based on highest SINR value\\
      \STATE Move UE on top of "overloaded" BS list to BS on top of "accepting" list\\
     \STATE Repeat for rest of UEs of first "overloaded" BS\\
      \STATE Repeat for rest of "overloaded" BSs\\
     \STATE \textbf{If} $N_{u} =\mu$ \textbf{then}\\ 
     \STATE Finalize UEs associated with that BS to it\\
      \STATE For moving or newly added UEs, calculate the new SINR matrix and repeat\\
      \STATE Repeat for the other tier (TeraHertz)\\
      \STATE Generate final load distribution per BS\\
     \end{algorithmic}
\end{algorithm}

\section{Numerical Results and Discussions}
Matlab$^{TM}$ simulation is conducted to analyze the performance of our proposed algorithms. The values for the parameters of our simulation are shown below:
\begin{itemize}
\item RF frequency=300 MHz and THz frequency=300 GHz
\item Total  number  of  TBSs  =76
\item Total  number  of  MBSs =10
\item No. of RF BS Antennas =1000
\item No. of THz BS Antennas =200
\item Working area = $2000\times2000 (m^{2})$
\item Min Allowed Distance Between RF BSs=400m
\item Min Allowed Distance Between THz BSs=100m
\item Path Loss Exponent  =3
\item Molecular absorption coefficient ($k_{a}(f))$=0.0016m$^{-1}$
\item SINR Threshold =0.5
\item Standard Deviation Threshold ($\epsilon$) = 1
\end{itemize}

\begin{figure}[t!]
\centering
\includegraphics[width=3.3in]{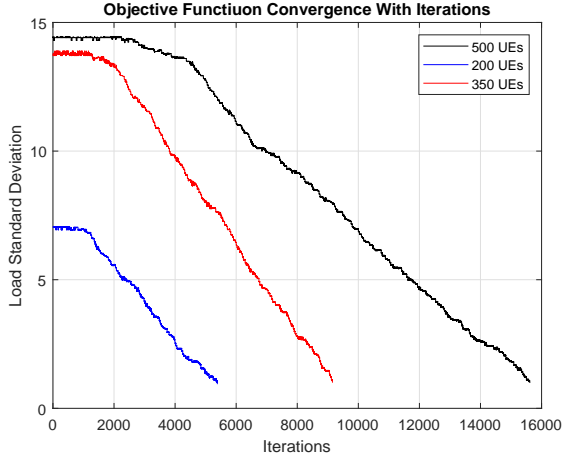}
\caption{Objective Function Convergence for LSTD Algorithm}
\label{u22}
\end{figure}

\begin{figure}[t!]
\centering
\includegraphics[width=3.3in]{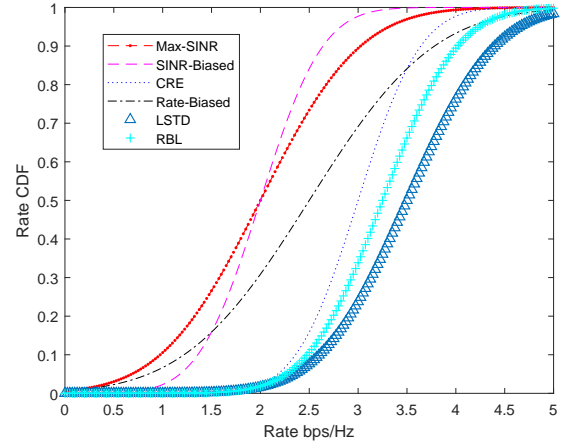}
\caption{Data Rate Comparison of Proposed Algorithms}
\label{u3}
\end{figure}

\begin{figure}[t!]
\centering
\includegraphics[width=3.3in]{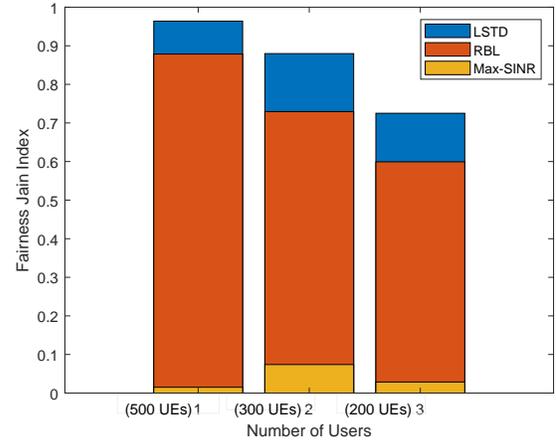}
\caption{Fairness Jain's Index Comparison of Proposed Algorithms}
\label{u4}
\end{figure}

Fig. \ref{u22} shows the standard deviation of the first proposed algorithm as it converges with iterations. It is obvious that as there are more UEs to be associated in the network, the objective function takes slightly more time to converge. Next, Fig.~\ref{u3} shows data rate of various algorithms compared with our proposed algorithms. For max-SINR scheme, most UEs select the MBSs so less resources are available to them. SINR-based scheme provides a slight improvement in data rate. CRE and rate-based scheme provide a significant improvement over max-SINR scheme, where more UEs are offloaded from MBSs to TBSs due to the biasing factor. As a result, more resources are available for MBS UEs.
Finally, Fig. \ref{u4} shows Jain's Index of our proposed algorithms compared to max-SINR scheme. 
It is interesting to note that fairness is improved with more UEs associated in the network for our proposed algorithms. As more UEs are available in the network, BSs get more opportunities to associate UEs to them and network load is balanced in a better way so that fairness is improved. LSTD and RBL achieve nearly the same performance with LSTD yielding a slight improvement over RBL (LSTD=0.96 and RBL=0.89 for 500 UEs).


\section{Conclusion and Future Research Directions}
In this paper, we analyzed the performance of a massive MIMO-enabled coexisting RF and THz network. We noted that conventional user association schemes may not be well-suited in 6G coexisting RF and THz networks. Subsequently, we proposed two user association algorithms using tools from unsupervised learning. Several unique challenges, however, have still to be addressed to achieve the full potential of THz communications. For instance, THz transmissions incur very high propagation losses, which significantly limit the communication distances. Hence, while in aerial, satellite, and vehicular networks, THz frequencies can provide low-latency communication, the propagation losses can hinder the gains. Furthermore, the coexistence of mmWave, sub 6GHz, and optical wireless communications and networking is not yet fully understood. Furthermore,  reconfigurable intelligent surfaces, ultra-massive MIMO configurations, and integrated access and backhaul, can boost the gains of THz communications. 

\bibliographystyle{IEEEtran}
\bibliography{IEEEabrv,pro}


\end{document}